\documentclass[a4paper,11pt]{article}
\pdfoutput=1
\usepackage{typearea}
\areaset[0cm]
        {16cm}{23cm}         
\usepackage[latin1]{inputenc}
\usepackage[T1]{fontenc}

\usepackage{amsmath}
\usepackage{mathrsfs}
\usepackage{amssymb}
\usepackage{mathcomp}
\usepackage{dsfont}
\usepackage{slashed}
\usepackage{mathtools}
\usepackage{cancel}
\usepackage{colortbl}
\usepackage{color}
\usepackage{accents}

\usepackage{graphicx}
\usepackage{float}
\usepackage[footnotesize]{caption}
\usepackage{subcaption}

\definecolor{head}{gray}{0.9}
\newcolumntype{A}{%
>{\columncolor{head}}l}

\newcommand*{\mathcolor}{}
\def\mathcolor#1#{\mathcoloraux{#1}}
\newcommand*{\mathcoloraux}[3]{%
  \protect\leavevmode
  \begingroup
    \color#1{#2}#3%
  \endgroup}

\newcommand{\SU}[2]{\text{SU}(#1)_{\text{#2}}}
\newcommand{\U}[2]{\text{U}(#1)_{\text{#2}}}

\newcommand{\betaKHeavy}[1]{(\accentset{\,(n+1) } { \beta}
^{\scriptscriptstyle{\quad\mathrm{heavy}=\SuperField{\nu}_{n},[1]}}
_{#1 \rightarrow #1 +\Delta #1^{\text{tree}}})_{\scriptscriptstyle{gf}}}

\newcommand{\betaQHeavy}[1]{(\beta_{Q\rightarrow Q+\Delta Q^{\mathrm{tree}}}^{\scriptscriptstyle{\lambda^{\{1\}},[1]}})_{i_1' \, \ldots \, i_n'}}

\newcommand{\Deltakappa}[1]{ \Delta\kappa_{\scriptscriptstyle{\SuperField{\nu}_{n}}}^{\scriptscriptstyle\text{#1}} }

\newcommand{\betaheavy}[2]{ \accentset{ (n+1) } {( \beta_{#1} }{}^{\scriptscriptstyle{\text{heavy}=\SuperField{\nu}_{ n},[1]}})_\sss{#2}}

\newcommand{\ssum}[2]{ \scalebox{0.75}{$\displaystyle\sum _{#1}^{#2}$} } 
\newcommand{\sss}[1]{ {\scriptscriptstyle{#1}} }
\newcommand{\bigpl}{ \raise3pt\hbox{$\displaystyle\Bigl($} }
\newcommand{\bigpr}{ \raise3pt\hbox{$\displaystyle\Bigr)$} }
\newcommand{\RaiseBrace}[1]{\raise3pt\hbox{$\displaystyle#1$}}
\newcommand{\SuperField}[1]{{#1}} 

\DeclareMathOperator{\Tr}{Tr}

\def\Nf{i}
\def\Ng{j}	
\def\ChargeC{\mathrm{C}}

\begin{document}

\begin{titlepage}

\vspace*{-15mm}
\vspace*{0.7cm}

\begin{center}

{\Large {\bf One-Loop Right-Handed Neutrino Threshold Corrections for \\[1mm] Two-Loop Running in Supersymmetric Type I Seesaw Models}}\\[8mm]

\vspace*{0.70cm}

Stefan Antusch$^{\star\dagger}$\footnote{Email: \texttt{stefan.antusch@unibas.ch}} and
Eros Cazzato$^{\star}$\footnote{Email: \texttt{e.cazzato@unibas.ch}}

\end{center}

\vspace*{0.20cm}

\centerline{$^{\star}$ \it
Department of Physics, University of Basel,}
\centerline{\it
Klingelbergstr.\ 82, CH-4056 Basel, Switzerland}

\vspace*{0.4cm}

\centerline{$^{\dagger}$ \it
Max-Planck-Institut f\"ur Physik (Werner-Heisenberg-Institut),}
\centerline{\it
F\"ohringer Ring 6, D-80805 M\"unchen, Germany}

\vspace*{1.2cm}

\begin{abstract}
\noindent The renormalization group (RG) running of the neutrino mass operator is required for comparing the predictions of neutrino models at high energies with the experimental data at low energies. In the type I seesaw scenario with $n_G$ right-handed neutrinos, the RG running is also performed in the effective theories above and between the thresholds given by the masses of the right-handed neutrinos. At these thresholds, the effective theories are matched. When calculating the two-loop RG running, the matching has to be performed at the one-loop level. In this work, we calculate the one-loop matching formulae in the MSSM extended by $n_G$ right-handed neutrinos using supergraph techniques. Moreover we present a general formula for one-loop matching of superpotential operators which can readily be applied to any supersymmetric theory where chiral superfields are integrated out.

\end{abstract}

\end{titlepage}

\newpage

\section{Introduction}

The origin of neutrino masses is one of the great open puzzles in particle physics. One of the best motivated mechanisms for generating the observed masses is the type I seesaw mechanism \cite{Minkowski:1977sc}, where $n_G$ right-handed neutrinos are added to the particle content of the Standard Model (SM). When the masses of the right-handed neutrinos are much larger than the electroweak scale (EW), this can explain the smallness of the masses of the light neutrinos (after EW symmetry breaking). The type I seesaw mechanism can also be embedded in extensions of the SM such as e.g.\ in Two-Higgs-Doublet Models or in the Minimal Supersymmetric Standard Model (MSSM). 

In order to compare the prediction of neutrino models, which are defined at high energy, with the experimental data obtained at low energies, one has to calculate the renormalization group (RG) running of the relevant quantities. Above the mass threshold of the heaviest of the right-handed neutrinos, these include in particular the neutrino Yukawa couplings and the mass matrix of the right-handed neutrinos. Below the mass threshold of the lightest of the right-handed neutrinos, the heavy particles are integrated out of the theory generating the effective dimension five neutrino mass operator, and its running has to be computed. Between the mass thresholds, one has to deal with the effective theories where the neutrino Yukawa matrix, the right-handed neutrino mass matrix as well as the neutrino mass operator are present. At the thresholds, the effective theories are matched. 

The renormalization group equations (RGEs) for the running of the neutrino mass operator have been calculated at one-loop in the SM \cite{Antusch:2001ck}, in Two-Higgs-Doublet Models \cite{Antusch:2001vn} and in the MSSM \cite{Chankowski:1993tx,Babu:1993qv,Antusch:2001vn}. For one-loop running tree-level matching is sufficient, and the formalism and RGEs for the intermediate effective theories have been described in \cite{Antusch:2002rr}. In the MSSM extended by $n_G$ right-handed neutrinos, the RGE for the running of the neutrino mass operator has been calculated at the two-loop level in \cite{Antusch:2002ek}. However, for consistent two-loop running one also needs to compute the matching of the effective theories at one-loop level, also referred to as the one-loop threshold corrections.

In this work, we calculate the one-loop matching formulae in the MSSM extended by $n_G$ right-handed neutrinos using supergraph techniques. Moreover we present a general formula for one-loop matching of superpotential operators which can be applied to any supersymmetric theory where chiral superfields are integrated and the effective theories are matched at the mass thresholds. We also comment on other choices of the matching scale, which may simplify the matching procedure in some cases.

The paper is organized as follows: In section 2 we review neutrino mass generation in the MSSM extended by $n_G$ right-handed neutrinos, and the effective theories which arise from integrating out the heavy particles at their mass thresholds. Section 3 contains a brief review of the method for calculating RGEs using supergraph techniques (following \cite{Antusch:2002ek}). The general formula for one-loop matching of superpotential operators in supersymmetric theories, when integrating out chiral superfields, is derived in section 4 and applied to the MSSM extended by $n_G$ right-handed neutrinos in section 5. Section 6 contains a summary and our conclusions.

\section{MSSM with Right-Handed Neutrinos: Integrating out and Effective Theories}
\label{sec:Model}

In order to take into account the observed neutrino masses in the Minimal Supersymmetric Standard Model (MSSM), we consider the MSSM extended by $n_G$ singlet superfields $\SuperField{\nu}^{\ChargeC j}$ ($j=1,\dots,n_G$), which contain right-handed neutrinos as fermionic components. When they have large (Majorana) masses, this provides an explanation for the smallness of the neutrino masses after electroweak  (EW) symmetry breaking via the (type I) seesaw mechanism  \cite{Minkowski:1977sc}.

The Yukawa part of the superpotential which includes the additional term with the neutrino Yukawa matrix $Y_\nu$, and the part of the superpotential with the mass matrix $M$ of the right-handed neutrino superfields, are given by
\begin{eqnarray} \label{eq:Superpotential Full} \nonumber  
 \mathscr{W}
 &=&
 (Y_e)_{gf}\SuperField{e}^{\ChargeC g}
 	\SuperField{h}^{(1)}_a\varepsilon^{ab}\SuperField{\ell}^f_b
 +(Y_d)_{gf}\SuperField{d}^{\ChargeC g}
 	\SuperField{h}^{(1)}_a\varepsilon^{ab}\SuperField{q}^f_b
 +(Y_u)_{gf}\SuperField{u}^{\ChargeC g}
 \SuperField{h}^{(2)}_a (\varepsilon^T)^{ab}  \SuperField{q}_b^f 
 \\&& 
 +(Y_\nu)_{\Nf f}\SuperField{\nu}^{\ChargeC \Nf} 
	\SuperField{h}^{(2)}_a (\varepsilon^T)^{ab}\SuperField{\ell}^f_b
 +\frac{1}{2}\SuperField{\nu}^{\ChargeC \Nf}(M)_{\Nf \Ng}\SuperField{\nu}^{\ChargeC \Ng}  \;,
\end{eqnarray}
where $\varepsilon$ is the totally antisymmetric tensor in two dimensions, $a,b\in\{1,2\}$ are $\SU{2}{}$ indices, $f,g\in\{1,2,3\}$ are flavour indices and the indices $i,j\in\{1,\ldots,n_G\}$ run over the number of right-handed neutrino superfields. 
The eigenstates of the mass matrix $M$, $\{\SuperField{\nu}^{\ChargeC 1},\dots,\SuperField{\nu}^{\ChargeC n_G}\}$, are labelled in such a way that $M_1 < M_2 < \dots < M_{n_G}$.\footnote{ We assume here that the mass spectrum is not degenerate. The generalization to a (partially) degenerate spectrum is straightforward.} 

\begin{figure}[H]
\centering
\includegraphics[scale=.85]{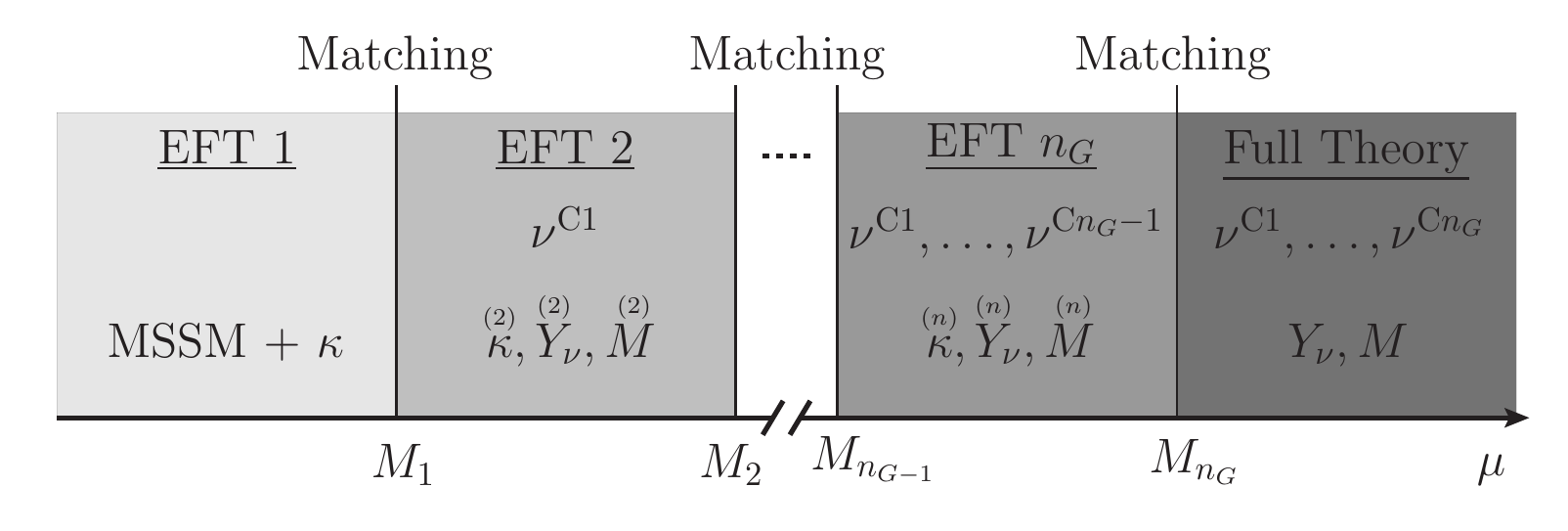}
\caption{\label{fig:EFT_succession} Illustration of the effective theories for the RG evolution in seesaw models with non-degenerate masses $M_{n}$ of the right-handed neutrinos. At the threshold $\mu = M_n$, the right-handed neutrino $\nu^{Cn}$ is integrated out of the theory and the ``EFT (n+1)'' and ``EFT n'' are matched. 
}
\end{figure}

In the following, we will consider an effective theory (EFT) description (see figure \ref{fig:EFT_succession}), using the same notation as in \cite{Antusch:2002rr}:
Above the highest mass threshold $M_{n_G}$, the ``Full Theory'' refers to the MSSM with all $n_G$ sterile neutrino superfields, which is described by the superpotential of eq.(\ref{eq:Superpotential Full}).  
At the threshold $M_{n_G}$, the heaviest of the right-handed neutrino superfields is integrated out, leading to the effective theory labelled ``EFT $n_G$'' which contains the effective dimension five neutrino mass operator and a reduced Yukawa matrix.  
We continue with this procedure and integrate out each sterile neutrino superfield $\SuperField{\nu}^{\ChargeC n}$ at the corresponding mass threshold $M_n$. 

Explicitly, for every intermediate region between the $(n\!-\!1)$th and the $n$th threshold, corresponding to ``EFT $n$'', the right-handed superfields $\{\SuperField{\nu}^{\ChargeC n},\dots,\SuperField{\nu}^{\ChargeC n_G}\}$ are integrated out, leading to the dimension five neutrino mass operator 
\begin{equation}\label{eq:Kappa-EFT}
 \mathscr{W}_{\kappa}^{\text{EFT } n} 
 =-\frac{1}{4} 
  \accentset{(n)}{\kappa}^{}_{gf} \, \SuperField{\ell}^{g}_c\varepsilon^{cd}
 \SuperField{h}^{(2)}_d\, 
 \, \SuperField{\ell}_{b}^{f}\varepsilon^{ba} \SuperField{h}^{(2)}_a  \;,
\end{equation}
with $\accentset{(n)}{\kappa}$ as the effective coupling matrix. In the region of the ``EFT $n$'', the Yukawa matrix for the remaining $n\!-\!1$ sterile neutrino superfields is reduced to a $(n\!-\!1)\times 3$ matrix which is referred to as $\accentset{(n)}{Y}_\nu$, i.e.\
\begin{equation}
 Y_\nu\;\longrightarrow\;
 \left(
  \begin{array}{ccc}
   	(Y_\nu)_{1,1} & (Y_\nu)_{1,2} & (Y_\nu)_{1,3} \\
    \vdots & \vdots & \vdots \\
 	(Y_\nu)_{n-1,1} & (Y_\nu)_{n-1,2} & (Y_\nu)_{n-1,3}\vspace*{1mm} \\
	\noalign{\hrule\vspace*{1mm}}
	0 & 0 & 0\cr
	\vdots & \vdots & \vdots \vspace*{1mm}\cr
	0 & 0 & 0 \cr
  \end{array}
 \right)
 \:
 \begin{array}{cl}
  \left.\begin{array}{c}\\[1.3cm]\end{array}\right\}
  & =: \accentset{(n)}{Y}_\nu\;,
  \\
  \left.\begin{array}{c}\\[1.3cm]\end{array}\right\}
  	&  \begin{array}{l}n_G\!-\!n\!+\!1\:\text{heavy, sterile}\\
  \text{neutrinos integrated out.}\end{array}
 \end{array}
\end{equation}
For each EFT it is convenient to denote the quark and charged lepton Yukawa matrices by $\accentset{(n)}{Y}_d $, $\accentset{(n)}{Y}_u $ and $\accentset{(n)}{Y}_e $. 
It is also useful to introduce $\accentset{(n)}{M}$ as the $(n\!-\!1)\times(n\!-\!1)$ matrix of the right-handed mass matrix below the $n$th threshold.
The superpotential of the ``EFT $n$'' now includes
\begin{eqnarray} \label{eq:Superpotential EFT} \nonumber  
 \mathscr{W}^{\text{EFT } n} 
 &=&
 (\accentset{(n)}{Y}_e)_{gf}\SuperField{e}^{\ChargeC g}
 	\SuperField{h}^{(1)}_a\varepsilon^{ab}\SuperField{\ell}^f_b
 +(\accentset{(n)}{Y}_d)_{gf}\SuperField{d}^{\ChargeC g}
 	\SuperField{h}^{(1)}_a\varepsilon^{ab}\SuperField{q}^f_b
 +(\accentset{(n)}{Y}_u)_{gf}\SuperField{u}^{\ChargeC g}
 \SuperField{h}^{(2)}_a (\varepsilon^T)^{ab}  \SuperField{q}_b^f 
 \\&& 
 +(\accentset{(n)}{Y}_\nu)_{\Nf f}\SuperField{\nu}^{\ChargeC \Nf} 
	\SuperField{h}^{(2)}_a (\varepsilon^T)^{ab}\SuperField{\ell}^f_b
 +\frac{1}{2}\SuperField{\nu}^{\ChargeC \Nf}(\accentset{(n)}{M})_{\Nf \Ng}\SuperField{\nu}^{\ChargeC \Ng}
   + \mathscr{W}_{\kappa}^{\text{EFT } n}  \;,
\end{eqnarray}
where the indices $i,j$ range from 1 to $n\!-\!1$.
Compared to the superpotential of the ``Full Theory'' the parameters of the effective superpotential now have a label ``$(n)$'', and the superpotential in addition contains the effective neutrino mass operator of eq.(\ref{eq:Kappa-EFT}).

At the $n$th threshold the tree-level matching condition for the the effective coupling constant reads
\begin{equation}
 \accentset{(n)}{\kappa}_\sss{gf} \Big|_{M_n} =  \accentset{\quad(n+1)\quad}{\quad\kappa\:_\sss{gf}} \Big|_{M_n}
 +2 \accentset{(n+1)}{(Y_{\nu}^T)}   _\sss{gn}  M^{-1}_n 
 	 \accentset{(n+1)}{(Y_{\nu})}  _\sss{nf} \Big|_{M_n}
 	 \qquad (\text{no sum over } n)\;,
\end{equation}
where $M_n$ corresponds to the largest eigenvalue of the $\accentset{(n+1)}{M}$ matrix.

The Yukawa matrices do not receive a threshold correction at tree-level
\begin{equation}
( \accentset{(n)}{Y}_{x} ) \Big|_{M_n} 
	=  \accentset{(n+1)}{(Y_{x})} \Big|_{M_n} ,
\end{equation}
where $x \;\in\; \{ d,u,e,\nu \}$. However, as we are going to discuss in section \ref{section5}, this will change at the one-loop level.

After successively integrating out all the right-handed neutrino superfields one arrives at the ``EFT 1'', corresponding to the MSSM with the dimension five neutrino mass operator 
\begin{equation}\label{eq:Kappa-MSSM}
 \mathscr{W}_{\kappa}^{\mathrm{MSSM}} 
 =-\frac{1}{4} 
  {\kappa}^{}_{gf} \, \SuperField{\ell}^{g}_c\varepsilon^{cd}
 \SuperField{h}^{(2)}_d\, 
 \, \SuperField{\ell}_{b}^{f}\varepsilon^{ba} \SuperField{h}^{(2)}_a \;,
\end{equation}
where one might drop the label (1) in $\accentset{(1)}{\kappa}$ and simply write $\kappa$. After EW symmetry breaking, $\kappa$ is related to the light neutrinos' mass matrix $m_\nu$ via
\begin{equation}\label{eq:Mnu-MSSM}
(m_\nu)_{gf} = \frac{1}{4}  {\kappa}_{gf} v_\mathrm{EW}^2\:,
\end{equation}
where $v_\mathrm{EW} \approx 246$ GeV.

\section{RGEs from Wave Function Renormalization Constants}
In this section, we review a formalism for computing $\beta$-functions for tensorial quantities of the superpotential from wave function renormalization constants, following \cite{Antusch:2001vn,Antusch:2002ek}. We will apply this formalism to establish a connection between the $\beta$-functions and the one-loop threshold corrections in section \ref{sec:1L-threshold corr}. We use modified dimensional reduction ($\overline{\text{DR}}$) \cite{Siegel:1979wq,Capper:1979ns} in $d=4-\epsilon$ dimensions.

\subsection{Derivation of the RGEs} \label{subsec:beta}

In order to compare high energy predictions for a (renormalized) quantity $Q$ with experimental results at low energies, one must evolve the predictions to low energies with the renormalization group equations (RGEs),
\begin{equation}
 \mu \frac{\mathrm d}{\mathrm d \mu} Q \; = \; \beta_Q \:,
\end{equation}
where $\mu$ is the renormalization scale and $\beta_Q$ the $\beta$-function.

We consider a general term of the superpotential expressed in bare quantities
\begin{equation}\label{eq:bare Q superpot. term}
(Q_B)_{i_1\,i_2\,\ldots\, i_n}(\Phi_B)_{i_1}(\Phi_B)_{i_2}\ldots(\Phi_B)_{i_n} \;,
\end{equation}
where the indices $i_x$ each specify a particular chiral superfield. Here, $n$ is the number of chiral superfields involved in the operator. For $n>3$, the superpotential operator is an effective operator. 

The superpotential term of eq.(\ref{eq:bare Q superpot. term}) can be recast in terms of renormalized quantities: 
\begin{align}\label{eq:bare & renormalized superpot. term}
(Q_B)_{i_1\,i_2\,\ldots\, i_n} \prod_{x=1}^n (\Phi_B)_{i_x}
\;=\;
Q_{i_1\,i_2\,\ldots\, i_n}\,\mu^{D_Q \epsilon}\;
\prod_{x=1}^n \Phi_{i_x}
\:.
\end{align}
$D_Q$ is related to the mass dimension of $Q$. The bare superfields are related to the renormalized ones by 
\begin{equation}\label{eq:renormalization superfields}
(\Phi_B)_{i_x} \;=\; Z_{i_x\,i'_x}^{1/2}\Phi_{i'_x} \, ,
\end{equation}
where $Z$ is the wave function renormalization constant, 
\begin{equation}
Z_{i_x\,i'_x} \;=\; \mathds{1}_{i_x\,i'_x}+\delta Z_{i_x\,i'_x} \:.
\end{equation}
Thus, inserting eq.(\ref{eq:renormalization superfields}) into eq.(\ref{eq:bare & renormalized superpot. term}), one obtains the relation of the bare quantity $Q_B$ to its renormalized counterpart $Q$:
\begin{equation}
(Q_B)_{i_1'\,i_2'\,\ldots\, i_n'}
\;=\;
Q_{i_1\,i_2\,\ldots\, i_n}\,\mu^{D_Q \epsilon}\;
Z^{-\frac{1}{2}}_{i_1 \, i_1'} Z^{-\frac{1}{2}}_{i_2\, i_2'}\ldots Z^{-\frac{1}{2}}_{i_n\,i_n'}
\;=\; 
Q_{i_1\,i_2\,\ldots\, i_n}\,\mu^{D_Q \epsilon}\;
\Bigg[ \prod_{x=1}^n Z^{-\frac{1}{2}}_{i_x \, i_x'} \Bigg]\:.
\end{equation}
Note the absence of vertex renormalization constants due to the non-renormalization theorem for supersymmetric theories \cite{Grisaru:1979wc}, which also holds for non-renormalizable operators \cite{Weinberg:1998uv}. 

The wave function renormalization constants depend on the renormalized variables of the theory, which we label as $\{v_{abc\ldots}\}$. For complex quantities $v_{abc\ldots}$ the complex conjugate variables $v_{abc\ldots}^*$ are treated as additional independent variables. The set $\{v_{abc\ldots}\}$ contains in particular the coupling $Q_{i_1\,\ldots\, i_n}$ (and $Q_{i_1\,\ldots\, i_n}^*$), but also the other couplings of the theory including e.g.\ the gauge couplings.

In the $\overline{\text{DR}}$ scheme, the wave function  renormalization constants can be expanded as
\begin{align}\label{eq:renormalization constant}
Z(\{ v_{abc\ldots} \})& \;=\; \mathds 1+\delta Z(\{ v_{abc\ldots} \}) \;=\; \mathds 1+ \sum_{k\geq 1} \frac{\delta Z_{,k}(\{ v_{abc\ldots} \})}{\hat\epsilon^k}\:,
\end{align}
where $\hat\epsilon$ is defined via
\begin{align}\label{eq:epsilon hat}
\frac{2}{\hat\epsilon}\; =\; \frac{2}{\epsilon} + \ln(4 \pi) - \gamma_E   \:,
\end{align}
with $d = 4 - \epsilon$ and where $ \gamma_E$ is the Euler-Mascheroni constant. In the following we will also use the notation 
\begin{align}\label{eq:DR bar}
 \Delta_{\overline{\text{DR}}}\; := \;\frac{2}{\hat\epsilon}\;=\;\frac{2}{\epsilon}+\ln (4\pi)-\gamma_E\:.
\end{align}
Notice that $v_{abc\ldots}(\mu)$ are functions of the renormalization scale $\mu$, whereas the bare quantities are per definition independent of $\mu$, and that the renormalization constants of eq.(\ref{eq:renormalization constant}) do not depend explicitly on $\mu$ (only implicitly via the $v_{abc\ldots}(\mu)$).

A detailed derivation for the calculation of the $\beta$-function from the wave function renormalization constants can be found in \cite{Antusch:2001vn,Antusch:2002ek}. There, the derivation was performed for minimal subtraction, however it also holds for $\overline{\text{DR}}$, with $\epsilon$ replaced by $\hat \epsilon$. 
The $\beta$-function for a quantity $Q$ (in $N=1$ supersymmetry) is given by 
\begin{equation}\label{eq:beta function master formula}
\beta_{Q}(\{v_{abc\ldots}\})_{i_1' \, \ldots \, i_n'}  \;=\; 
-\frac{1}{2} Q_{i_1 \, \ldots \, i_n}\sum_{x=1}^n \Bigg(\sum_{v_{abc\ldots}^{(*)}} D_{v_{abc\ldots}^{(*)}} \frac{\mathrm d (\delta Z)_{i_x \, i_x' \,,1}}{\mathrm d v_{abc\ldots}^{(*)}}v_{abc\ldots}^{(*)}\Bigg)  \prod_{y \neq x}\delta_{i_y i_y'} \:.
\end{equation}
We note that at the one- and two-loop level, the coefficients $\delta Z_{,k}(\{ v_{abc\ldots} \})$ of the wave function renormalization constants are identical in the $\overline{\text{DR}}$ and DR schemes (in fact in all mass independent schemes), as one can easily verify using e.g.\ the results of \cite{West:1984dg}. This also implies that in supersymmetric theories the two-loop $\beta$ functions for superpotential operators are the same  in the $\overline{\text{DR}}$ and DR schemes. In particular, the results of \cite{Antusch:2002ek} also hold in the $\overline{\text{DR}}$ scheme.

\subsection{Two-Loop RGEs in the MSSM with Right-Handed Neutrinos}\label{sec:rges}

The one-loop $\beta$-functions for the quantities $\accentset{(n)}{\kappa}$, $\accentset{(n)}{Y}_\nu$ and $\accentset{(n)}{M}$ of the EFTs in the MSSM with right-handed neutrinos can be found in \cite{Antusch:2002rr}. The complete results for the one and two-loop $\beta$-functions for the EFTs can be obtained using the wave function renormalization constants given in  \cite{Antusch:2002ek}, by adding the label ``(n)'' above each coupling to match our notation.\footnote{The two-loop RGEs for the gauge couplings including right-handed neutrinos, and a discussion of the effects of the right-handed neutrinos on gauge coupling unification, can be found e.g.\ in \cite{Casas:2000pa,Martin:1993zk}. We note that integrating out the right-handed neutrino superfield does not induce one-loop threshold corrections for the gauge couplings.}

The one-loop $\beta$-function for $\accentset{(n)}{\kappa}$ (with the number in square brackets indicating the loop order) is given by: 
\begin{eqnarray}\label{eq:RGEForKappaInEFTn}
16\pi^2 \accentset{(n)}{\beta}_\kappa^\sss{\;[1]} & = & 
 \bigpl\accentset{(n)}{Y}_e^\dagger \accentset{(n)}{Y}_e\bigpr^T \: \accentset{(n)}{\kappa}
 + \accentset{(n)}{\kappa} \, \bigpl\accentset{(n)}{Y}_e^\dagger \accentset{(n)}{Y}_e\bigpr
 + \bigpl \accentset{(n)}{Y}^\dagger_\nu  \accentset{(n)}{Y}_\nu \bigpr^T \,\accentset{(n)}{\kappa}
 + \accentset{(n)}{\kappa} \: \bigpl
 \accentset{(n)}{Y}^\dagger_\nu\accentset{(n)}{Y}_\nu\bigpr
\nonumber \\
&& {} + 2\, \Tr \bigpl \accentset{(n)}{Y}^{\dagger}_\nu 
 \accentset{(n)}{Y}_\nu\bigpr\,\accentset{(n)}{\kappa}
 +6\,\Tr\bigpl \accentset{(n)}{Y}_u^\dagger \accentset{(n)}{Y}_u\bigpr \,\accentset{(n)}{\kappa} 
 -\tfrac{6}{5} \,\accentset{(n)}{g_1}^{2} \;\accentset{(n)}{\kappa}- 6 \accentset{(n)}{g_2}^2 \;\accentset{(n)}{\kappa}
 \;.
\end{eqnarray}
For the Yukawa matrices below the $n$th threshold, the $\beta$-functions are 
\begin{eqnarray}\label{eq:RGEForYdInEFTn}
16\pi^2 \accentset{(n)}{\beta}_{Y_d}^\sss{\;[1]}
 &=& 
 \accentset{(n)}{Y}_d
 \left[ 3\, \accentset{(n)}{Y}^\dagger_d \accentset{(n)}{Y}_d 
+ \accentset{(n)}{Y}_u^\dagger \accentset{(n)}{Y}_u
+ \Tr \bigpl\accentset{(n)}{Y}^{\dagger}_e \accentset{(n)}{Y}_e \bigpr 
+3\Tr \bigpl\accentset{(n)}{Y}_d^\dagger \accentset{(n)}{Y}_d \bigpr
- \tfrac{7}{15}\accentset{(n)}{g_1}^2 - 3 \accentset{(n)}{g_2}^2 -\tfrac{16}{3}\accentset{(n)}{g_3}^2 \right] \!\! \;,
\\
16\pi^2 \accentset{(n)}{\beta}_{Y_u}^\sss{\;[1]}
 &=& 
 \accentset{(n)}{Y}_u
 \left[ 3\, \accentset{(n)}{Y}^\dagger_u\accentset{(n)}{Y}_u 
+ \accentset{(n)}{Y}_d^\dagger \accentset{(n)}{Y}_d
+ \Tr \bigpl\accentset{(n)}{Y}^{\dagger}_\nu\accentset{(n)}{Y}_\nu \bigpr 
+3\Tr \bigpl\accentset{(n)}{Y}_u^\dagger \accentset{(n)}{Y}_u \bigpr
- \tfrac{13}{15}\accentset{(n)}{g_1}^2 - 3 \accentset{(n)}{g_2}^2 -\tfrac{16}{3}\accentset{(n)}{g_3}^2 \right]\!\!  \;,
\\
16\pi^2 \accentset{(n)}{\beta}_{Y_e}^\sss{\;[1]}
 &=& 
 \accentset{(n)}{Y}_e
 \left[ 3\, \accentset{(n)}{Y}^\dagger_e\accentset{(n)}{Y}_e 
+ \accentset{(n)}{Y}_\nu^\dagger \accentset{(n)}{Y}_\nu
+ \Tr \bigpl\accentset{(n)}{Y}^{\dagger}_u\accentset{(n)}{Y}_u \bigpr 
+3\Tr \bigpl\accentset{(n)}{Y}_d^\dagger \accentset{(n)}{Y}_d \bigpr 
- \tfrac{9}{5}\accentset{(n)}{g_1}^2 - 3 \accentset{(n)}{g_2}^2 \right] \!\! \;,
\\ \label{eq:RGEForYnuInEFTn}
16\pi^2 \accentset{(n)}{\beta}_{Y_\nu}^\sss{\;[1]}
 &=& 
 \accentset{(n)}{Y}_\nu
 \left[ 3\, \accentset{(n)}{Y}^\dagger_\nu\accentset{(n)}{Y}_\nu 
+ \accentset{(n)}{Y}_e^\dagger \accentset{(n)}{Y}_e
+ \Tr \bigpl\accentset{(n)}{Y}^{\dagger}_\nu  \accentset{(n)}{Y}_\nu \bigpr 
+3\Tr \bigpl\accentset{(n)}{Y}_u^\dagger \accentset{(n)}{Y}_u \bigpr
- \tfrac{3}{5}\accentset{(n)}{g_1}^2 - 3 \accentset{(n)}{g_2}^2 \right] \!\! \;,
\end{eqnarray}
and for the Majorana mass matrix of the right-handed neutrinos the one-loop $\beta$-function reads
\begin{eqnarray}\label{eq:RGEForMBetweenThresholdsMSSM}
16\pi^2 \accentset{(n)}{\beta}_{M}^\sss{\;[1]}
&=& 
   2\, \bigpl \accentset{(n)}{Y}_\nu \accentset{(n)}{Y}^\dagger_\nu \bigpr \, \accentset{(n)}{M} 
   + 2\,\accentset{(n)}{M}\,\bigpl \accentset{(n)}{Y}_\nu  \accentset{(n)}{Y}^\dagger_\nu \bigpr^T \;.
\end{eqnarray}
Note that we used the GUT charge normalization for the $\U{1}{Y}$ charge. We list  the results for the one-loop RGEs here explicitly since we will use them for the one-loop threshold corrections.

\section{One-Loop Threshold Corrections for Superpotential Operators}\label{section4}
\label{sec:1L-threshold corr}

For consistent two-loop running within mass-independent renormalization schemes, one has to take into account one-loop threshold corrections from decoupling of heavy particles. The aim of this section is to derive a formalism for computing these one-loop threshold corrections using supergraphs. For our analysis we use again $\overline{\text{DR}}$ as renormalization scheme and choose to integrate out the heavy particles at their mass thresholds.\footnote{We will also comment on other choices of the matching scale.} We will focus on the case of integrating out heavy chiral superfields, since we later want to apply the formalism to the right-handed neutrinos.

\subsection{General Framework} \label{subsec:framework}

\begin{figure}[H]
\centering
\includegraphics[scale=.85]{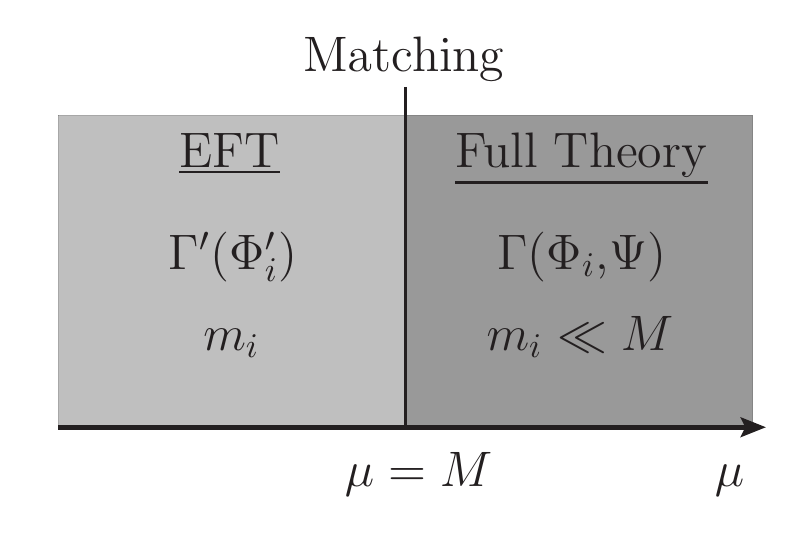}
\caption{\label{fig:EFT/FT} Illustration of the matching between full and effective theory when a chiral superfield $\Psi$ with mass $M$ is integrated out at its mass threshold. $\Gamma(\Phi_i,\Psi)$ is the effective action superfunctional of the full theory and $\Gamma'(\Phi'_i)$ the effective action superfunctional of the EFT. The superfunctionals are matched at $\mu = M$. 
The matching includes the canonical normalization of the fields $\Phi_i$, which are then called $\Phi'_i$ (as explained in the main text). 
}
\end{figure}

The general scenario we consider in this section is illustrated in figure \ref{fig:EFT/FT}. $\Phi_i$ represent light chiral superfields with masses $m_i$ and $\Psi$ stands for a heavy chiral superfield with mass $M$ (where $M \gg m_i$) which is integrated out at its mass threshold (i.e.\ at $\mu = M$). Above $M$, we refer to the theory as the ``Full Theory'' and below $M$ as the effective theory (EFT).

With $N_\Phi$ light chiral superfields $\Phi_i$ ($i\in \{1,\ldots, N_\Phi\}$) the general trilinear superpotential couplings (of the full theory) can be written as:
\begin{align}\label{eq:yukawas definition}
\mathscr W_\lambda
=  \frac{\lambda^{\{0\}}_{ijk}}{3!} \Phi_i\Phi_j\Phi_k 
 + \frac{\lambda^{\{1\}}_{ij}}{2!} \Psi\Phi_i\Phi_j
 + \frac{\lambda^{\{2\}}_{i}}{2!} \Psi\Psi\Phi_i
 + \frac{\lambda^{\{3\}}}{3!} \Psi\Psi\Psi \;,
\end{align}
where the the label in curly brackets specifies the number of heavy particles $\Psi$ coupling to each $\lambda$. Diagrammatically, we represent the heavy chiral superfield $\Psi$ by a straight double line while the light chiral superfields $\Phi_i$ are represented by straight single lines. The supergraph diagrams corresponding to the  trilinear couplings of eq.(\ref{eq:yukawas definition}) are shown in figure \ref{fig:Yuks}:
\begin{figure}[H]
\centering
$\vcenter{\hbox{\includegraphics[scale=0.85]{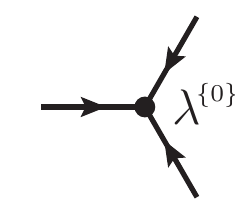}}}\;, 
\vcenter{\hbox{\includegraphics[scale=0.85]{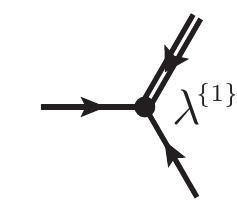}}}\;, 
\vcenter{\hbox{\includegraphics[scale=0.85]{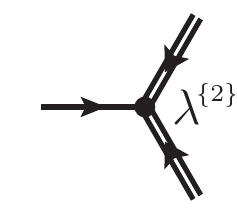}}}\;, 
\vcenter{\hbox{\includegraphics[scale=0.85]{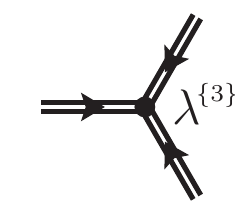}}} \:.$
\caption{\label{fig:Yuks}Diagrammatic representation of the trilinear couplings.}
\end{figure}

\subsection{One-Loop Threshold Correction for a Quantity $\boldsymbol{Q}$} \label{subsec:quantityQ}

We now consider the one-loop threshold correction for a general quantity $Q$ corresponding to a superpotential operator

\begin{align} \label{eq:Q-operator}
\mathscr W_Q
= Q_{i_1 \, \ldots \, i_n} \prod_{x=1}^n \Phi_{i_x} \:.
\end{align}
Note that for $n=3$, this is just the term corresponding to $\lambda^{\{0\}}_{ijk} \Phi_i\Phi_j\Phi_k $. And for $n>3$, $Q$ is understood as an effective operator of the superpotential. 

\subsubsection*{Strategy}
Due to the non-renormalization theorem such superpotential operators will not receive any loop corrections, however they will in general be modified indirectly via the loop corrections to the two-point vertex functions proportional to $\Phi_i^\dagger \Phi_j$. These corrections change the normalization of the superfields. Canonically normalizing them involves superfield transformations $\Phi_i \to \Phi'_i$ which then implies a modification of the operator coefficient $Q \to Q'$. This is exactly the one-loop threshold correction which we want to compute.

Let's look at the matching of the two-point vertex function explicitly, when a heavy chiral superfield with mass $M$ is integrated out at its mass threshold. Including supergraph one-particle irreducible (OPI) diagrams up to one-loop order, we obtain in both theories, the full theory and the EFT:\footnote{Note that effective operators,  which may be present in the superpotential, do not contribute to the one-loop matching of the two-point vertex function.} 
\begin{figure}[H]
\begin{subfigure}{\textwidth}
$\Gamma\hphantom{'} \;\supset\;
\vcenter{\hbox{\includegraphics[scale=0.45]{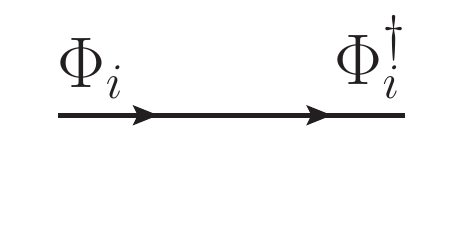}}}
+
\vcenter{\hbox{\includegraphics[scale=0.45]{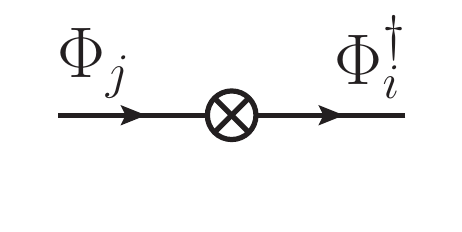}}}
+
\vcenter{\hbox{\includegraphics[scale=0.45]{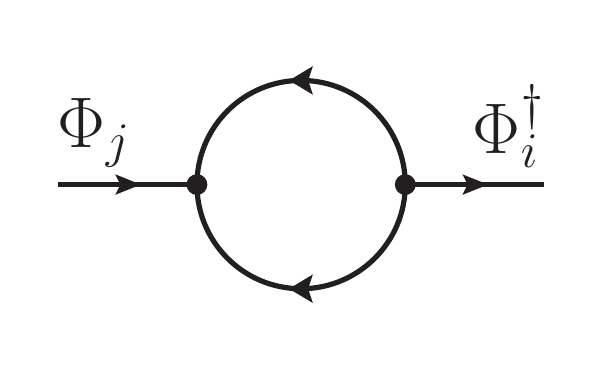}}}
+
\vcenter{\hbox{\includegraphics[scale=0.45]{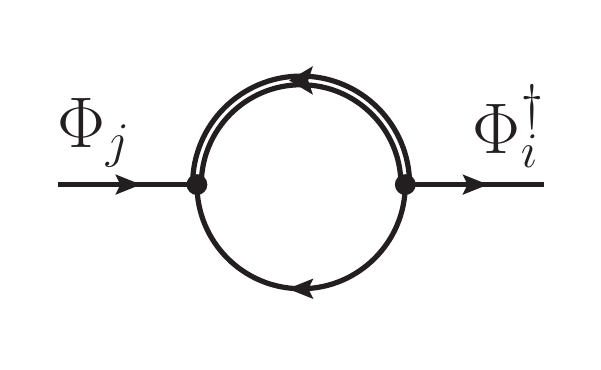}}}
+
\vcenter{\hbox{\includegraphics[scale=0.45]{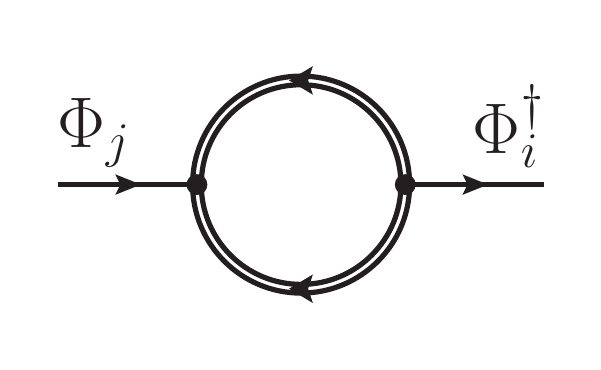}}}$
\caption{\label{fig:2-point eff action full}Relevant OPI diagrams of the one-loop effective action of the full theory.}
\end{subfigure}
\begin{subfigure}{\textwidth}
$\Gamma' \;\supset\;
\vcenter{\hbox{\includegraphics[scale=0.45]{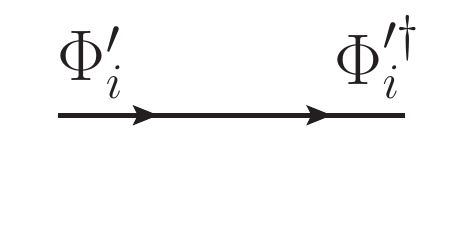}}}
+
\vcenter{\hbox{\includegraphics[scale=0.45]{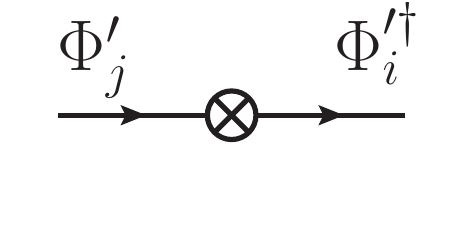}}}
+
\vcenter{\hbox{\includegraphics[scale=0.45]{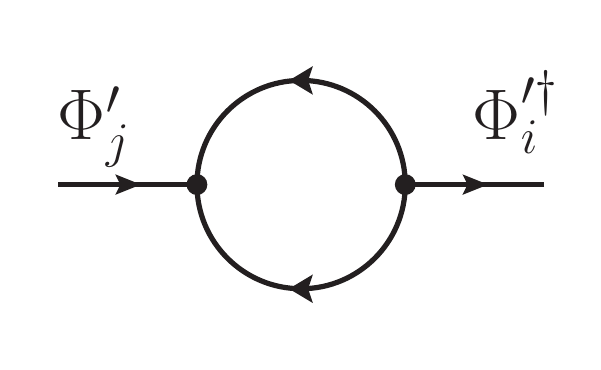}}}$
\caption{\label{fig:2-point eff action EFTT}Relevant OPI diagrams of the one-loop effective action of the EFT.}
\end{subfigure}
\caption{\label{fig:2-point eff action}Part of the one-loop effective action which contains the two-point function for the chiral superfields. The crosses are the one-loop counterterms. }
\end{figure}
Note that in the EFT we use the fields $\Phi'_i$, which are understood to be canonically normalized. The condition for this canonical normalization, as well as the effective operator content of the EFT, are calculated from the matching condition 
\begin{align}\label{eq:matching}
\Gamma^{'}(\Phi'_i) \;\stackrel{\mu = M}{=}\;\Gamma(\Phi_i,\xcancel{\Psi}) \:.
\end{align}
The crossed $\Psi$ indicates that the heavy field is integrated out of the theory.

\subsubsection*{Integrating Out Heavy Chiral Superfields at One-Loop}
Let us now explicitly consider the integrating out of a heavy chiral superfield $\Psi$ at the one-loop level. As discussed above, we can focus on the two-point vertex function and integrate out the heavy internal particles from the loop diagrams in figure \ref{fig:2-point eff action full}.

Considering the one-loop supergraph of figure \ref{fig:2-point eff action full} with one internal heavy field in the limit $p^2,m^2\ll M^2$ gives:
\begin{align}  \nonumber  
\vcenter{\hbox{\includegraphics[scale=0.55]{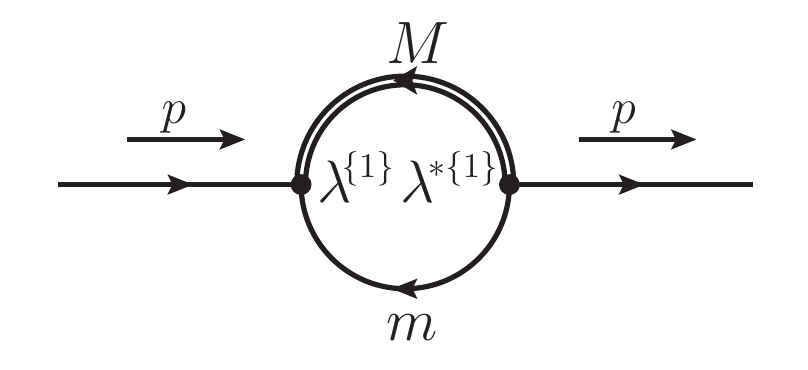}}}
=\;
&\int \mathrm d^4 \theta \int \frac{\mathrm d^4p}{(2\pi)^4} \;\Phi_i^\dag(-p,\bar \theta)\Phi_j(p,\theta)\frac{1}{16\pi^2}\lambda_{ik}^{*\{1\}}\lambda_{kj}^{\{1\}} B_0(p^2,m^2,M^2)
\\[-5pt] \nonumber
\xrightarrow{ p^2,m^2 \ll M^2} 
\int \mathrm d^4 \theta &\int \frac{\mathrm d^4p}{(2\pi)^4} \;\Phi_i^\dag(-p,\bar \theta)\Phi_j(p,\theta)\frac{1}{16\pi^2}\lambda_{ik}^{*\{1\}}\lambda_{kj}^{\{1\}}
\left(\Delta_{\overline{\text{DR}}}	- \ln \left( \frac{M^2}{\mu^2} \right) +1 \right)	
\\[5pt] \label{eq:1-loop 1xheavy}
\stackrel{\mu = M}{=} 
\int \mathrm d^4 \theta &\int \frac{\mathrm d^4p}{(2\pi)^4} \;\Phi_i^\dag(-p,\bar \theta)\Phi_j(p,\theta)\frac{1}{16\pi^2}\lambda_{ik}^{*\{1\}}\lambda_{kj}^{\{1\}}
\left(\Delta_{\overline{\text{DR}}}	 +1 \right)	\:.  
\end{align}
$B_0$ is one of the Passarino-Veltman functions. In the last line we inserted $\mu =M$, which corresponds to performing the matching at the mass threshold of the heavy field. 

The analogous steps are done for the second one-loop supergraph diagram with two internal heavy fields:
\begin{align}   \nonumber 
\vcenter{\hbox{\includegraphics[scale=0.55]{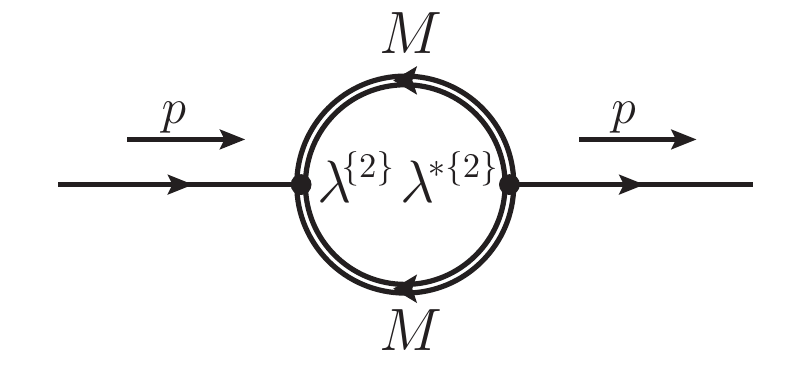}}}
=\; 
&\int \mathrm d^4 \theta \int \frac{\mathrm d^4p}{(2\pi)^4} \;\Phi_i^\dag(-p,\bar \theta)\Phi_j(p,\theta)\frac{1}{16\pi^2}\lambda_{i}^{*\{2\}}\lambda_{j}^{\{2\}} B_0(p^2,M^2,M^2)
\\[-5pt] \nonumber
\xrightarrow{ p^2 \ll M^2} 
\int \mathrm d^4 \theta &\int \frac{\mathrm d^4p}{(2\pi)^4} \;\Phi_i^\dag(-p,\bar \theta)\Phi_j(p,\theta)\frac{1}{16\pi^2}\lambda_{i}^{*\{2\}}\lambda_{j}^{\{2\}}
\left(\Delta_{\overline{\text{DR}}}	- \ln \left( \frac{M^2}{\mu^2} \right)  \right)	
\\[5pt] \label{eq:1-loop 2xheavy}
\stackrel{\mu = M}{=} 
\int \mathrm d^4 \theta &\int \frac{\mathrm d^4p}{(2\pi)^4} \;\Phi_i^\dag(-p,\bar \theta)\Phi_j(p,\theta)\frac{1}{16\pi^2}\lambda_{i}^{*\{2\}}\lambda_{j}^{\{2\}}
\left(\Delta_{\overline{\text{DR}}}	+0  \right)	\;.
\end{align}
Note that for $\mu =M$ the whole expression from this diagram is cancelled by the counterterm (in the $\overline{\text{DR}}$ scheme), leaving no finite part which contributes to the matching.

\subsubsection*{Comment on the choice of the matching scale}

So far, we have focused on $\mu =M$, however it may sometimes be desirable to choose a different matching scale, in particular when this leads to a simplification of the matching procedure.  
For instance, for the choice $\mu =M/\sqrt{e}$, the expression from the diagram in eq.(\ref{eq:1-loop 1xheavy}) is completely cancelled by the counterterm. Then, however,  the diagram of eq.(\ref{eq:1-loop 2xheavy}) contributes to the matching proportional to $-  \ln \left( {M^2}/{\mu^2} \right)  = -1$. The generalisation of our treatment to different choices of the matching scale is straightforward. We will come back to this possibility after eq.(\ref{eq:matching 1-loop}) and in section \ref{section5}. For the remainder of this section, we will again focus on the case $\mu =M$.

\subsubsection*{Matching and Canonical Normalization}

We now perform the matching according to eq.(\ref{eq:matching}) (cf.\ figure \ref{fig:2-point eff action}). This will require a field redefinition (i.e.\ a canonical renormalization) for which we can make the following general ansatz:  
\begin{subequations}\label{eq:Phi transf}
\begin{align} 
 \Phi'_i \;=\;& \big(\delta_{ij} + \tfrac{1}{2} (\Delta \Phi)_{ij}\big) \Phi_j \qquad\qquad\quad\hphantom{^{\dag}}\longrightarrow&
 \Phi _j &\;=\; \big(\delta_{ji} - \tfrac{1}{2} (\Delta \Phi)_{ji}\big) \Phi'_i\;,
\\
 \Phi'^\dag_i =\;&\Phi_j^\dag  \big(\delta_{ji} + \tfrac{1}{2} (\Delta \Phi^\dag)_{ji}\big)\qquad\qquad\quad \longrightarrow&
 \Phi^\dag_j &\;=\; \Phi'^\dag_i \big(\delta_{ij} - \tfrac{1}{2} (\Delta \Phi^\dag)_{ij}\big) \;,
\end{align}
\end{subequations}
where $\Delta \Phi$ and $\Delta \Phi^\dag$ are regarded as small quantities. Comparing the diagrams above and below the threshold, we obtain (at the given one-loop order):
\begin{align}\label{eq:matching 2-point fct}
\int \mathrm d^8z\;\Phi'^\dag_i \, \delta_{ij} \,\Phi'_j \;\stackrel{!}{=}\; 
\int \mathrm d^8z\;
\Phi^\dag_i \bigg( \delta_{ij} +  \frac{1}{16\pi^2}\lambda_{ik}^{*\{1\}}\lambda_{kj}^{\{1\}} \bigg) \Phi_j \:.
\end{align}

Note that in figure \ref{fig:2-point eff action} there are loop diagrams with light internal fields which exist both above and below the thresholds. However since the diagrams are loop-suppressed and since $\Delta \Phi$ is a small quantity they differ only at the level of small quantities squared, and thus drop out at the considered order. 

Inserting the ansatz for the field redefinition into eq.(\ref{eq:matching 2-point fct}), and expanding up to first order in the small quantities, we obtain
\begin{align} \nonumber
\Phi'^\dag_i \, \delta_{ij} \,\Phi'_j &\;\stackrel{!}{=}\; 
\Phi^\dag_i \bigg( \delta_{ij} +  \frac{1}{16\pi^2}\lambda_{ik}^{*\{1\}}\lambda_{kj}^{\{1\}} \bigg) \Phi_j
\\ \nonumber
&\;=\; \Phi'^\dag_{i'} \big(\delta_{i'i} - \tfrac{1}{2} (\Delta \Phi^\dag)_{i'i}\big)  
\bigg( \delta_{ij} +  \frac{1}{16\pi^2}\lambda_{ik}^{*\{1\}}\lambda_{kj}^{\{1\}} \bigg) 
\big(\delta_{jj'} - \tfrac{1}{2} (\Delta \Phi)_{jj'}\big) \Phi'_{j'}
\\ 
&\;=\; \Phi'^\dag_{i'} \bigg(\delta_{i'j'} - \tfrac{1}{2} (\Delta \Phi^\dag)_{i'j'} - \tfrac{1}{2} (\Delta \Phi)_{i'j'}
+ \frac{1}{16\pi^2}\lambda_{i'k}^{*\{1\}}\lambda_{kj'}^{\{1\}} +\ldots \bigg)   \Phi'_{j'} \:. \label{eq:2 pint matching}
\end{align}
The dots in the last line represent second and higher order terms. From eq.(\ref{eq:2 pint matching}) we can conclude
\begin{align}
 (\Delta \Phi)_{ij} \;=\; 
\frac{1}{16\pi^2}\lambda_{ik}^{*\{1\}}\lambda_{kj}^{\{1\}} \:. 
\end{align}
$\Delta\Phi$ is solely determined by the one-loop supergraph of eq.(\ref{eq:1-loop 1xheavy}).

\subsubsection*{One-Loop Matching for a Quantity $\boldsymbol{Q}$}
Let us now turn to the matching of a quantity $Q$ which corresponds to a superpotential operator as defined in eq.(\ref{eq:Q-operator}). Due to the non-renormalization theorem the matching condition (at $\mu=M$) reads
\begin{align}
Q'_{i_1 \, \ldots \, i_n} \prod_{x=1}^n  \Phi'_{i_x} \;\stackrel{!}{=}\;
(Q+\Delta Q^\sss{\text{tree}})_{i_1 \, \ldots \, i_n} \prod_{x=1}^n \Phi_{i_x} \:.
\end{align}
$\Delta Q^\sss{\text{tree}}$ stands for a contribution to the n-point vertex function with heavy internal fields which contributes to the (effective) operator below the the mass threshold when the heavy fields get integrated out. For example, when $Q$ corresponds to the neutrino mass operator, then $\Delta Q^\sss{\text{tree}}$ is a contribution to it from integrating out a heavy right-handed neutrino with mass $M$ at $\mu=M$, as discussed in section \ref{sec:Model}. $Q'$ is the quantity of the effective theory below the threshold. The one-loop corrections enter via the canonical normalization $\Phi_i \to \Phi_i'$.

Inserting eq.(\ref{eq:Phi transf}) and expanding to first order in $\Delta \Phi$, one finds
\begin{align} \nonumber
Q'_{i_1 \, \ldots \, i_n} \prod_{x=1}^n  \Phi'_{i_x} \;\stackrel{!}{=}\;
&(Q+\Delta Q^\sss{\text{tree}})_{i_1 \, \ldots \, i_n} \prod_{x=1}^n \Phi_{i_x}
\\  \nonumber
\stackrel{(\ref{eq:Phi transf})}{=}
&(Q+\Delta Q^\sss{\text{tree}})_{i_1 \, \ldots \, i_n} 
\prod_{x=1}^n  \big(\delta_{i_x  i'_x} - \tfrac{1}{2} (\Delta \Phi)_{i_x\,i'_x}\big) \Phi'_{i'_x} 
\\ 
\;=\;\;
&(Q+\Delta Q^\sss{\text{tree}})_{i_1 \, \ldots \, i_n} 
\bigg( \prod_{x=1}^n  \delta_{i_x  i'_x} \Phi'_{i'_x} 
- \frac{1}{2} \sum_{x=1}^n (\Delta \Phi)_{i_x\,i'_x}  \Phi'_{i'_x} 
\prod_{y \neq x} \delta_{i_y  i'_y}\Phi'_{i'_y}  +\ldots \bigg)\:.
\end{align}
From this relation, we can extract the one-loop threshold correction to $Q$ (at $\mu=M$):
\begin{align}\label{eq:Q'=Q+..}
 Q'_{i'_1 \, \ldots \, i'_n} 
\;=\; (Q+\Delta Q^\sss{\text{tree}})_{i'_1 \, \ldots \, i'_n} 
- \frac{1}{2}  (Q+\Delta Q^\sss{\text{tree}})_{i_1 \, \ldots \, i_n}\sum_{x=1}^n (\Delta \Phi)_{i_x\,i'_x}  
\prod_{y \neq x} \delta_{i_y  i'_y}\:.
\end{align}

\subsubsection*{Relation to the $\boldsymbol{\beta}$ function}
As shown above, $\Delta\Phi$ is solely determined by the one-loop supergraph diagram of eq.(\ref{eq:1-loop 1xheavy}). Since the $1/\hat\epsilon$ part and the finite part are related (in the $\overline{\text{DR}}$ scheme) we can express $\Delta\Phi$ in terms of the corresponding part of $\delta Z$, which we label $\delta Z^\sss{\lambda^{\{1\}},[1]}$. The couplings of type $\lambda^{\{1\}}$ are defined in eq.(\ref{eq:yukawas definition}) and ``one'' in square brackets indicates the loop order.   
We obtain 
\begin{align}\label{eq:Z lambda 1}
(\delta Z^\sss{\lambda^{\{1\}},[1]})_{ij} \;=\; -\frac{1}{16\pi^2}\lambda_{ik}^{*\{1\}}\lambda_{kj}^{\{1\}} \Delta_{\overline{\text{DR}}} 
\qquad \xRightarrow{\mathrm{eq}.(\ref{eq:DR bar})} \qquad
(\delta Z_{,1}^\sss{\lambda^{\{1\}},[1]})_{ij} \;=\; -\frac{2}{16\pi^2}\lambda_{ik}^{*\{1\}}\lambda_{kj}^{\{1\}}\;,
\end{align}
which implies that $\Delta \Phi$ is related to $\delta Z^\sss{\lambda^{\{1\}},[1]}$ via
\begin{align} \label{eq:Delta Phi = d Z}
(\Delta \Phi)_{ij} \;=\; 
\frac{1}{16\pi^2}\lambda_{ik}^{*\{1\}}\lambda_{kj}^{\{1\}} 
\;=\;
-\frac{1}{2}(\delta Z_{,1}^\sss{\lambda^{\{1\}},[1]})_{ij} \:.
\end{align}
The part of $\beta_Q$ which is solely determined by $\delta Z^{\lambda^{\{1\}},[1]}$ can now be written as (in the ${\overline{\text{DR}}}$ scheme):
\begin{align} \nonumber \label{eq:Beta Q of Z lambda 1}
(\beta_{Q}^\sss{\lambda^{\{1\}},[1]})_{i_1' \, \ldots \, i_n'}  &\stackrel{(\ref{eq:beta function master formula})}{=}
-\frac{1}{2} Q_{i_1 \, \ldots \, i_n}\sum_{x=1}^n \Bigg(\sum_{v_{abc\ldots}^{(*)}} D_{v_{abc\ldots}^{(*)}} \frac{\mathrm d (\delta Z_{,1}^\sss{\lambda^{\{1\}},[1]})_{i_x \, i_x'} }{\mathrm d v_{abc\ldots}^{(*)}}v_{abc\ldots}^{(*)}\Bigg)  \prod_{y \neq x}\delta_{i_y i_y'}
\\ \nonumber
 & \stackrel{\phantom{(\ref{eq:beta function master formula})}}{=}
-\frac{1}{2} Q_{i_1 \, \ldots \, i_n}\sum_{x=1}^n (\delta Z_{,1}^\sss{\lambda^{\{1\}},[1]})_{i_x \, i_x'} \prod_{y \neq x}\delta_{i_y i_y'}
\\
&\stackrel{(\ref{eq:Delta Phi = d Z})}{=}
  Q_{i_1 \, \ldots \, i_n}\sum_{x=1}^n (\Delta \Phi)_{i_x\,i'_x}  
\prod_{y \neq x} \delta_{i_y  i'_y}\:.
\end{align}
In the first line, the derivative is taken with respect to all quantities of the theory $v_{abc\ldots}$, and we inserted $D_{\lambda^{\{1\}}}=\frac{1}{2}$ for the $\varepsilon$-dependence of the trilinear couplings. Note that as usual $v_{abc\ldots}$ and $v_{abc\ldots}^*$ are treated as independent variables.

Using eq.(\ref{eq:Beta Q of Z lambda 1}), we can relate the second term on the r.h.s. of eq.(\ref{eq:Q'=Q+..}) to the part of $\beta_Q$ from $\delta Z^\sss{\lambda^{\{1\}},[1]}$ and thus rewrite the one-loop threshold correction to $Q$ at $\mu =M$ as:
\begin{align} \label{eq:matching 1-loop}
 Q'_{i'_1 \, \ldots \, i'_n} 
\;=\; (Q+\Delta Q^\sss{\text{tree}})_{i'_1 \, \ldots \, i'_n} 
- \frac{1}{2}\betaQHeavy{Q}
\:.
\end{align}
Notice that in the $\beta$-function the $Q$ is replaced by $Q+\Delta Q^\sss{\text{tree}}$, which is indicated by the label $Q \to Q+\Delta Q^\sss{\text{tree}}$.\footnote{\label{fn7} We note that for the alternative choice $\mu =M/\sqrt{e}$ mentioned above, the diagram in eq.(\ref{eq:1-loop 2xheavy}) contributes instead of the diagram in 
eq.(\ref{eq:1-loop 1xheavy}). This implies that in eq.(\ref{eq:matching 1-loop}) the expression $\betaQHeavy{Q}$ has to be replaced by $-(\beta_{Q\rightarrow Q+\Delta Q^{\mathrm{tree}}}^{\scriptscriptstyle{\lambda^{\{2\}},[1]}})_{i_1' \, \ldots \, i_n'}$.}

\section{Application to Right-Handed Neutrino Thresholds}\label{section5}

In this section we apply the above-derived formalism to the model described in section \ref{sec:Model}, i.e.\ to the MSSM extended by $n_G$ right-handed neutrino superfields. We compute the one-loop threshold corrections for the running of the effective coupling matrix $\kappa$ of the neutrino mass operator, the Yukawa coupling matrices and the right-handed neutrino mass matrix.

\subsection*{One-Loop Matching of the Neutrino Mass Operator}

As described in section \ref{sec:Model}, the threshold corrections are applied at $\mu = M_n$, when the corresponding right-handed neutrino with mass $M_n$ is integrated out and the ``EFT $n+1$'' and ``EFT n'' are matched.\footnote{\label{fn8}
We note that, as discussed in section \ref{section4}, other choices of the matching scale are possible as well. In the specific case of right-handed neutrino thresholds, since there are no trilinear vertices with two right-handed neutrino superfields, one may choose to match at $\mu = M_n/\sqrt{e}$. Then, the contribution from the diagram in eq.(\ref{eq:1-loop 1xheavy}) is cancelled by the counterterm, and there is no contribution as in eq.(\ref{eq:1-loop 2xheavy}) due to the absence of the corresponding vertex. This means that for the specific choice $\mu = M_n/\sqrt{e}$, instead of a shift in the quantities $Q$, the one-loop threshold correction is accounted for by the rescaling of the matching scale. For the remainder of this section, we will discuss the case $\mu =M_n$.
} The one-loop matching condition reads (using eq.(\ref{eq:matching 1-loop})) 
\begin{equation} \label{eq:1LmatchingKappa}
 \accentset{(n)}{\kappa}_\sss{gf} \Big|_{M_n} =  \accentset{\quad(n+1)\quad}{\quad\kappa\:_\sss{gf}} \Big|_{M_n}
 +( \Deltakappa{tree}) _\sss{gf} \Big|_{M_n}
 +( \Deltakappa{loop}) _\sss{gf} \Big|_{M_n}\:,
\end{equation}
where $\Deltakappa{tree}$ is the tree-level correction and $\Deltakappa{loop}$ the one-loop correction, given by
\begin{eqnarray}
( \Deltakappa{tree}) _\sss{gf} \Big|_{M_n} &=& 
2 \accentset{(n+1)}{(Y_{\nu}^T)}   _\sss{gn}  M^{-1}_n 
 	 \accentset{(n+1)}{(Y_{\nu})}  _\sss{nf} \Big|_{M_n}
 	 \qquad (\text{no sum over } n)\:,\\
 ( \Deltakappa{loop}) _\sss{gf} \Big|_{M_n} &=&
 -\frac{1}{2} \betaKHeavy{\kappa}\Big|_{M_n}
  \:.	 	 
\end{eqnarray}
On the left side of the equations, the subscript $\SuperField{\nu}_{n}$ indicates that the right-handed neutrino superfield with mass eigenvalue $M_n$ is integrated out of the theory. Similarly, on the right side of the second equation, ``heavy=$\nu_n$'' indicates that $\SuperField{\nu}_{n}$ is the heavy superfield which gets integrated out at $\mu = M_n$.

In the MSSM extended by $n_G$ right-handed neutrino superfields, the coupling $\lambda^{ \{1\} }$, introduced in the previous section, can be identified with the $n$th row of the neutrino Yukawa matrix contained in $\accentset{(n+1)}{Y_{\nu} }$, i.e.\ with $\accentset{(n+1)}{(Y_{\nu}) }_\sss{ng}$ ($n$ fixed, $g$ runs from 1 to 3).
From the one-loop $\beta$-function in eq.(\ref{eq:RGEForKappaInEFTn}), we thus obtain:
\begin{align}
-  \frac{1}{2}  \betaKHeavy{\kappa} \Big|_{M_n}
= - \frac{1}{32 \pi^2 } 
&\left[
\ssum{h=1}{3}\raisebox{1pt}{$\bigl($}\;\;\accentset{(n+1)}{\kappa}\;+\Deltakappa{tree}\raisebox{1pt}{$\bigr)$}_\sss{gh} \accentset{(n+1)}{(Y_{\nu}^\dag)} _\sss{hn} \accentset{(n+1)}{(Y_{\nu}) }   _\sss{nf}
\right. \nonumber
\\ \nonumber
& \left.	
	+ 
	 \ssum{h=1}{3}\bigpl \accentset{(n+1)}{(Y_{\nu}^\dag)}   _\sss{hn} \accentset{(n+1)}{(Y_{\nu}) }   _\sss{ng}\bigpr^T 		\raisebox{1pt}{$\bigl($}\;\;\accentset{(n+1)}{\kappa}\;+\Deltakappa{tree}\raisebox{1pt}{$\bigr)$}_\sss{hf}  
\right.  \nonumber
\\
&\left.
	+ 2 \Tr \bigpl \accentset{(n+1)}{(Y_{\nu}^\dag)} _\sss{hn} \accentset{(n+1)}{(Y_{\nu}) }  _\sss{nl} \bigpr
	 \raisebox{1pt}{$\bigl($}\;\;\accentset{(n+1)}{\kappa}\;+\Deltakappa{tree}\raisebox{1pt}{$\bigr)$}_\sss{gf} 
  \right]_{M_n} \:.
  \end{align}
  
\subsection*{One-Loop Matching of the Neutrino Yukawa Matrix}

The neutrino Yukawa matrix does not receive a threshold correction at tree-level, i.e.\
\begin{equation}
( \accentset{(n)}{Y}_{\nu} ) _\sss{ig} \Big|_{M_n} 
	=  \accentset{(n+1)}{(Y_{\nu})}  _\sss{ig} \Big|_{M_n} \:,
\end{equation}
where the index $i$ runs from 1 to $n-1$ and $g$ from 1 to 3. Extending the matching to the one-loop order, we get
\begin{equation}\label{eq:1LmatchingYnu}
( \accentset{(n)}{Y}_{\nu} ) _\sss{ig}  \Big|_{M_n} 
	=   \accentset{(n+1)}{(Y_{\nu})}  _\sss{ig} \Big|_{M_n} 
	- \frac{1}{2} \betaheavy{Y_{\nu}}{ig} \Big|_{M_n} ,
\end{equation}
with the one-loop threshold correction given by (no sum over $n$)
\begin{align}
- \frac{1}{2} \betaheavy{Y_{\nu}}{ig} \Big|_{M_n}
= -  \frac{1}{32 \pi^2 } 
&\left[
	3 \,\ssum{f=1}{3}  \accentset{(n+1)}{(Y_{\nu})} _\sss{if}  \accentset{(n+1)}{(Y_{\nu}^\dag)}  _\sss{fn}  \accentset{(n+1)}{(Y_{\nu})}  _\sss{ng}
	+
	 \accentset{(n+1)}{(Y_{\nu})}  _\sss{ig} \Tr \bigpl  \accentset{(n+1)}{(Y_{\nu}^\dag)}  _\sss{fn} \accentset{(n+1)}{(Y_{\nu})} _\sss{nh}\bigpr 
\right]_{M_n} \:.
\end{align}

\subsection*{One-Loop Matching of the Right-Handed Neutrino Mass Matrix}

The one-loop matching condition of the right-handed neutrino mass matrix at $\mu = M_n$ is given by
\begin{equation}\label{eq:1LmatchingM}
\accentset{(n)}{(M)}_\sss{ij} \Big|_{M_n} = \accentset{(n+1)}{(M)}_\sss{ij} \Big|_{M_n}
	- \frac{1}{2}\betaheavy{M}{ij} \Big|_{M_n} ,
\end{equation}
with the one-loop threshold correction equal to (no sum over $n$)
\begin{align}
-  \frac{1}{2}  \betaheavy{M}{ij} \Big|_{M_n} =
	-\frac{1}{16 \pi^2 }\left[
	\ssum{f=1}{3}  \accentset{(n+1)}{(Y_{\nu})} _\sss{if}  \accentset{(n+1)}{(Y_{\nu}^\dag)}  _\sss{fn}  \accentset{(n+1)}{(M)}_\sss{nj}
	+\ssum{f=1}{3} \accentset{(n+1)}{(M)}_\sss{in}  \accentset{(n+1)}{(Y_{\nu}^*)}  _\sss{nf} \accentset{(n+1)}{(Y_{\nu}^T)}  _\sss{fj} 
	\right]_{M_n} \:.
\end{align}
We note that at the threshold $\mu = M_n$, in order to integrate out a mass eigenstate, we go to the mass eigenbasis of the right-handed neutrinos, i.e.\ we diagonalize the matrix $\accentset{(n+1)}{M}$. However, the matrix $\accentset{(n)}{M}$ is not necessarily diagonal at $\mu = M_n$ due to the threshold correction.

\subsection*{One-Loop Matching of the Quark and Charged Lepton Yukawa Matrices}
The one-loop matching condition for the Yukawa matrices of the quarks and charged leptons is given by
\begin{equation} \label{eq:1LmatchingYx}
( \accentset{(n)}{Y}_{x} ) _\sss{gf} \Big|_{M_n} 
	=   \accentset{(n+1)}{(Y_{x})}  _\sss{gf} \Big|_{M_n} 
	- \frac{1}{2} \betaheavy{Y_x}{gf} \Big|_{M_n} \:,
\end{equation}
where $x\;\in\; \{  d,u,e \}$. The one-loop threshold corrections are given by
\begin{eqnarray}
-  \frac{1}{2}  \betaheavy{Y_d}{gf} \Big|_{M_n} &=& 0  \;,
\\
-  \frac{1}{2}  \betaheavy{Y_u}{gf} \Big|_{M_n} &=& - \frac{1}{32 \pi^2 } 
\left[(\accentset{(n+1)}{Y_u})_\sss{gf} \Tr \bigpl \accentset{(n+1)}{(Y_{\nu}^\dag)} _\sss{hn} \accentset{(n+1)}{(Y_{\nu}) }  _\sss{nl} \bigpr \right]_{M_n} \quad(\text{no sum over $n$})\;,
\\
-  \frac{1}{2}  \betaheavy{Y_e}{gf} \Big|_{M_n} &=& - \frac{1}{32 \pi^2 } 
\left[\ssum{h=1}{3}(\accentset{(n+1)}{Y_e})_\sss{gh} \accentset{(n+1)}{(Y_{\nu}^\dag)} _\sss{hn} \accentset{(n+1)}{(Y_{\nu}) }   _\sss{nf} \right]_{M_n} \quad(\text{no sum over $n$})\;.
\end{eqnarray}

\section{Summary and Conclusions}

In this paper, we have derived a general formula (cf.\ eq.(\ref{eq:matching 1-loop})) for the one-loop matching of superpotential operators using supergraph techniques, which can readily be applied to any supersymmetric theory where chiral superfields are integrated out and the effective theories are matched at the mass thresholds.\footnote{We also discussed other choices of the matching scale (cf.\ subsection below eq.(\ref{eq:1-loop 2xheavy}) and footnote \ref{fn7}), which can lead to a simplification of the matching procedure (cf.\ footnote \ref{fn8}).} 
We have applied our formula to calculate the one-loop threshold corrections in the MSSM extended by $n_G$ right-handed neutrinos. 
These results (cf.\ section 5) can now be used to study the running in a type I seesaw extension of the MSSM consistently at two-loop. The procedure can be summarized as follows:
\begin{itemize}

\item From a ``Full Theory'' at high energies, the quantities are evolved using the relevant two-loop RGEs (cf.\ section \ref{sec:rges}) down to the first mass threshold, corresponding to the heaviest sterile neutrino superfield with the largest eigenvalue $M_{n_G}$ of the mass matrix $M$. 

\item  At the threshold  $\mu = M_{n_G}$ the heaviest sterile neutrino superfield is integrated out and the one-loop matching conditions are imposed according to eq.(\ref{eq:1LmatchingKappa}), (\ref{eq:1LmatchingYnu}), (\ref{eq:1LmatchingM}) and (\ref{eq:1LmatchingYx}). To integrate out $\nu_{n_G}$ it is necessary to diagonalize the right-handed neutrino mass matrix by an unitary transformation $U$, $M\;\rightarrow\; U^T\,M\,U$, which corresponds to transforming the right-handed neutrino superfields by $\SuperField{\nu}^{\ChargeC}\;\rightarrow\;U^\dag \,\SuperField{\nu}^{\ChargeC}$. 
This also implies that the neutrino Yukawa matrix is transformed to $Y_\nu\;\rightarrow\; U^T \,Y_\nu$. 

\item Below the threshold, at $\mu < M_{n_G}$, we are in the effective theory referred to as ``EFT $n_G$'', where $n_G\!-\!1$ right-handed neutrino superfields are left.
Within this effective theory, the parameters (including $\accentset{(n_G)}{Y_\nu}\,$, $\accentset{(n_G)}{M}$, $\,\accentset{(n_G)}{Y_d}$, $\,\accentset{(n_G)}{Y_u}$, $\,\accentset{(n_G)}{Y_e}$ and the new effective coupling $\,\accentset{(n_G)}{\kappa}\,\,\,$) are evolved according to their RGEs down to the next threshold, corresponding to the largest eigenvalue of the $(n_G\!-\!1)\times(n_G\!-\!1)$ matrix $\accentset{(n_G)}{M}$, i.e. to $\mu = M_{n_G-1}$. 

\item At the threshold $\mu = M_{n_G-1}$, we repeat the steps of the second bullet point and integrate out $\nu_{n_G-1}$ at the one-loop level, leading to the ``EFT $n_G -1$''. This procedure is repeated up to the ``EFT 1'' where all right-handed neutrino superfields are integrated out.

\end{itemize}

This procedure yields in particular the low energy values of the superpotential couplings at the two-loop level. The one-loop threshold corrections from integrating out the heavy right-handed neutrinos affect directly the low energy values of the light neutrino mass matrix $m_\nu$ as well as of the up-type quark Yukawa matrix $Y_u$ and the charged lepton Yukawa matrix $Y_e$ (and various other quantities indirectly via the coupled RGEs). These corrections are, for instance, relevant for testing more precisely the predictions of supersymmetric Grand Unified Theories and flavour models.

\section*{Acknowledgements}
We thank Vinzenz Maurer and Constantin Sluka for helpful discussions. This work was supported by the Swiss National Science Foundation.


\providecommand{\href}[2]{#2}\begingroup\raggedright\endgroup

\end{document}